\documentclass[aps,twocolumn,pra,%
preprintnumbers,amsmath,amssymb,superscriptaddress,%
nofootinbib]{revtex4}
\usepackage{graphicx}
\usepackage{color}
\newcommand{\bea}{\begin{eqnarray}}
\newcommand{\eea}{\end{eqnarray}}
\newcommand{\p}{\partial}
\usepackage{bm}
\bmdefine{\bma}{\bm{a}}
\bmdefine{\bmOmega}{\bm{\Omega}}
\definecolor{orange}{rgb}{1,0.5,0}
\newcommand{\calG}{\mathcal{G}}

\newcommand{\calO}{\mathcal{O}}
\newcommand{\calW}{\mathcal{W}}
\newcommand{\vecq}{\Vec{q}}

\newcommand{\vecY}{\Vec{Y}}
\newcommand{\hkappa}{\Hat{\kappa}}
\newcommand{\hmu}{\Hat{\mu}}
\newcommand{\homega}{\Hat{\omega}}
\newcommand{\hOmega}{\Hat{\Omega}}
\newcommand{\Exp}[1]{\left\langle #1\,\right\rangle}

\newcommand{\retG}{\calG}

\newcommand{\band}{\mathcal{B}}
\newcommand{\tQNM}{\text{QNM}}

\begin{document}

\title{Holographic superfluid flows with a localized repulsive potential}


\author{Akihiro {\sc Ishibashi}}\email[]{akihiro@phys.kindai.ac.jp}
\affiliation{%
{\it Department of Physics, Kindai University, Higashi-Osaka 577-8502, JAPAN 
}}

\author{Kengo {\sc Maeda}}\email[]{maeda302@sic.shibaura-it.ac.jp}
\affiliation{%
{\it Faculty of Engineering,
Shibaura Institute of Technology, Saitama 330-8570, JAPAN
}}

\author{Takashi {\sc Okamura}}\email[]{tokamura@kwansei.ac.jp}
\affiliation{%
{\it Department of Physics,
Kwansei Gakuin University,
Sanda, Hyogo, 669-1337, JAPAN}}

\begin{abstract}
We investigate a holographic model of superfluid flows with an external 
repulsive potential. When the strength of the potential is sufficiently weak, 
we analytically construct two steady superfluid flow solutions. 
As the strength of the potential is increased, the two solutions merge into 
a single critical solution at a critical strength, and then disappear above 
the critical value, as predicted by a saddle-node bifurcation theory. 
We also analyze the spectral function of fluctuations around the solutions 
under a certain decoupling approximation. 
\end{abstract} 

\maketitle


\noindent

\section{Introduction}\label{sec:intro}

Superfluid flow in Bose-Einstein condensate in cold atoms or in ${}^4$He 
is of particular interest as an example of inviscid fluid flows with 
no dissipation. It is typically unstable due to the excitation of 
the flow (the Landau instability) or against the creation of solitons 
such as vortices (the soliton-emission instability). 
In particular, the latter instability is expected to occur in spatially 
inhomogeneous systems or around an obstacle~\cite{Feynman1955}.

The soliton-emission instability in inhomogeneous systems has been 
extensively investigated by solving the time-dependent Gross-Pitaevskii~(GP) equation~\cite{Gross1961, Pitaevskii1961}. 
Frisch et.~al.~\cite{Frisch1992} numerically found a steady superfluid 
flow solution in a two-dimensional system 
and showed that the flow is broken by the creation of vortices 
beyond a threshold velocity. 
Hakim~\cite{Hakim1997} obtained two analytic steady superfluid 
flow solutions in one-dimensional system in the presence of an external localized repulsive potential. 
They merge at a critical velocity, and beyond it, there is no steady flow solution. These features are consistent with experimental results~\cite{Raman1999, Onofrio2000, Inouye2001}, 
and they can be described in terms of a saddle-node bifurcation of 
the stationary solutions~\cite{PomeauRica1993}. 
However we should note that although the GP equation is very useful 
to understand various features of superfluid flows, it is applicable only to 
weakly interacting low temperature systems.

One of the possible approaches to tackling strongly interacting cases 
is to appeal to a ``holographic model'' based on 
the AdS/CFT duality~\cite{Maldacena:1997re}. 
There have already been a number of works on holographic models~\cite{HHH1, HHH2} in which strongly correlated condensed matter systems can be 
successfully described in terms of some gravitational theory 
via the AdS/CFT duality (see also Ref.~\cite{Horowitz_Review} for a review). 

In this paper, we shall construct a holographic superfluid model 
as an attempt to extend the analysis of the one-dimensional superfluid flow 
in~\cite{Hakim1997} to the strongly coupled case.  
We perturbatively construct analytic superfluid flow 
solutions~(see also \cite{HSW2011, IIM2014} for similar flow solutions) 
in the presence of an external localized repulsive potential. 
We find that two steady flow solutions exist below a critical value of the strength of the potential. One of the solutions is unstable 
and steeper near the potential than the stable one, 
as shown in \cite{Hakim1997}. 
The two solutions merge into one at the critical value and 
then disappear beyond it, as predicted by 
a saddle-node bifurcation theory.

We also derive the spectral function $\rho$ of fluctuations of 
the solution and investigate the effects of the impurity 
generated by the repulsive potential, under a certain decoupling approximation.
We find that with respect to a small variation of the chemical potential, 
the spectral function shows a peculiar behavior which is explained by the existence of a band gap generated 
by the external localized repulsive potential. 
  
This paper is organized as follows.  
In Sec.~\ref{sec:II}, we set up our holographic model that corresponds 
to the one-dimensional superfluid flow studied in~\cite{Hakim1997}. 
In Sec.~\ref{sec:III}, we perturbatively construct two steady solutions 
of holographic superfluid flows and derive the superfluid current. 
In Sec.~\ref{sec:IV}, we shall show that the steeper solution is 
unstable in the sense that the free energy is higher than 
that of the other solution.  
We study the spectral function in Sec.~\ref{sec:V}. 
Sec.~\ref{sec:VI} is devoted to summary and discussions.

\section{Holographic superfluid model}
\label{sec:II}
We will construct a holographic superfluid model which is dual to a strongly coupled field theory in $3+1$-dimensional 
Minkowski spacetime with the action 
\begin{align}
\label{action_probe}
& S=\int d^5 x\sqrt{-g}\left[R+\frac{12}{L^2}+\frac{L^2}{e^2}{\cal L}_m \right]+S_{\text{ct}},  \nonumber \\ 
& {\cal L}_m := - | D \psi|^2
  - \left( m^2 + V(x,u) \right) |\psi|^2
  - \frac{1}{4}F_{\mu\nu}F^{\mu\nu}~,
\end{align} 
where $\psi$ is a complex scalar field with mass $m$ and charge $e$, 
$D_\mu := \nabla_\mu - i\, A_\mu$, and where $L$ denotes 
the AdS radius, $V(x,u)$ a localized external repulsive 
potential, and $S_{\text{ct}}$ the counter term defined below.

In this paper, we consider a probe limit $e\to \infty$ in which the gauge field 
$A_\mu$ and the scalar field $\psi$ do not backreact on the original metric. 
Therefore we consider, as our background spacetime, the Schwarzschild-AdS 
metric with the temperature $T$: 
\begin{align}
\label{Schwarz_AdS}
  & \frac{ds^2}{L^2}
  = \frac{\pi^2 T^2}{u}\, (- f\, dt^2 + dx^2 + dy^2 + dz^2)
  +\frac{du^2}{4u^2 f}, 
\end{align}
where $f(u) := 1 - u^2$, and
$0<u<1$ outside the black hole with AdS boundary at $u=0$.

The field equations on our background are 
\begin{align}
\label{eq:scalar}
& D^\mu D_\mu\psi-m^2\psi-V(x,u)\psi=0 \,, \\
\label{eq:gauge}
& \nabla_\nu F^{\nu\mu}=i[\psi^\ast D^\mu\psi-\psi(D^\mu \psi)^\ast]
\,. 
\end{align}

For simplicity, we shall set $m^2L^2=-4$ so that the 
Breitenlohner-Freedman(BF) bound~\cite{BFbound} is saturated, 
and consider a periodic potential with a dimensionless positive constant $\hat{g}$
\begin{align}
\label{external_potential}
  V(x,u)=\frac{\hat{g}}{L}\,u\sum_{n=-\infty}^\infty\delta(x-nl) \,, 
\end{align}
localized at $x_n=nl~(n\in \mathbb{N})$. 
Note that $V$ corresponds to a repulsive potential, as $({\hat{g}}/{L})\,u$ is positive.

Since $V(x,u)$ decreases to zero as $u \rightarrow 0$ and thus 
the effective mass-squared saturates the BF bound at the AdS boundary,  
the asymptotic form of $\psi$ becomes 
\begin{align}
\label{asym_scalar}
\psi(u,x)\simeq - \alpha(x) u+\beta(x) u\ln u. 
\end{align} 
According to the dictionary of AdS/CFT duality, 
the coefficient $\alpha$ corresponds to the expectation value $\Exp{{\cal O}}$ 
of the dual field theory operator of dimension two, 
while $\beta$ corresponds to a source term in the dual boundary 
field theory~\cite{Herzog:2010}. 
Hereafter, we shall impose $\beta=0$ at the AdS boundary, 
as our boundary condition on $\psi$. 

As indicated in \cite{HSW2011}, it is convenient to use the following gauge invariant variables $R$ and $M_\mu$ defined by 
$\psi=R e^{i\varphi}/L$~(where $R$ is a dimensionless quantity) and $M_\mu:=A_\mu-\nabla_\mu \varphi$. 
Then, Eqs.~(\ref{eq:scalar}) and (\ref{eq:gauge}) reduce to the equations 
only for the gauge-invariant variables $R$ and $M_\mu$, 
\begin{align}
\label{eq:Real}
\nabla^2R-M^\mu M_\mu R-m^2R-V(x,u)R=0\,, \\
\label{eq:Imaginary}
\nabla^\mu(M_\mu R^2)=0 \,, \\
\label{eq:Maxwell}
\nabla_\nu F^{\nu\mu}=\frac{2R^2}{L^2}M^\mu \,, 
\end{align}
where Eqs.~(\ref{eq:Real}) and (\ref{eq:Imaginary}) are derived 
from the real and imaginary parts of Eq.~(\ref{eq:scalar}), respectively. 
When the system is stationary, Eq.~(\ref{eq:Imaginary}) implies 
the conservation of momentum of the superfluid flow.

According to the Bloch's theorem~\cite{Bloch}, 
having the periodic potential $V$, Eq.~(\ref{eq:scalar}) must admit 
a solution $\psi$ that is periodic in $x$, except the phase $\varphi$.  
This implies that for such a periodic solution $\psi$, the corresponding 
gauge invariants, $R$ and $M_\mu$, must also be periodic in $x$. 
So, hereafter we shall regard $R$ and $M_\mu$ as smooth functions 
(apart from the location of the delta function 
in Eq.~(\ref{external_potential})) 
on an annulus $(u,x) \in [0,1] \times {\bf S}^1$ 
with $x=0$ and $x=l$ being identified.

Since the potential is $x$-dependent, the velocity of the superfluid is 
also $x$-dependent. 
We consider the case that the superfluid velocity is injected 
at $x=l/2$
~(compare with \cite{Hakim1997}, 
in which the injection is made at spatial infinity). 
We impose the following asymptotic boundary conditions 
\begin{align} 
\label{asymt_M}
  & M_t(0,x)=\mu~,
& & M_x(0, l/2) = v_0~,
\end{align}
where $\mu$ and $v_0$ are respectively interpreted 
as the chemical potential and the superfluid velocity injected 
at $x= l/2$~(see, for example \cite{HKS2009}).

\section{Perturbative construction of the solutions, $(R, M)$}
\label{sec:III}
In this section we solve Eqs.~(\ref{eq:Real}), (\ref{eq:Imaginary}), 
and (\ref{eq:Maxwell}) perturbatively, assuming that the amplitude of $R$ 
be very small. Following \cite{MaedaOkamura2008}, 
we expand $R$ and $M_\mu$ as a series in a small parameter $\epsilon$ as
\begin{align}
\label{expansion}
& R=\epsilon^{\frac{1}{2}} R_1(u,x)+\epsilon^\frac{3}{2} R_2(u,x)+\cdots, \nonumber \\
& M_\mu=M_\mu^{(0)}(u,x)+\epsilon M_\mu^{(1)}(u,x)+\cdots, \nonumber \\
& F_{\mu\nu}=F^{(0)}_{\mu\nu}+\epsilon F^{(1)}_{\mu\nu}+\cdots, 
\end{align} 
where $F^{(i)}_{\mu\nu}=\p_\mu M^{(i)}_{\nu}-\p_\nu M^{(i)}_{\mu}$. 

At zeroth order in $\epsilon$, imposing that the chemical potential $M_t^{(0)}(0,x)$ is constant, 
we find the solution for $M_\mu^{(0)}(u,x)$ as 
\begin{align}
\label{sol:M0_5}
  & M_t^{(0)}=\mu_0(1-u)~,
& & M_x^{(0)}=\xi(x)~,
& & M_u^{(0)}=0~.
\end{align}
Here, $\xi(x)$ is the velocity of the superfluid satisfying $\xi(l/2)=v_0$ 
(see the condition~(\ref{asymt_M})).

Under the ansatz $R_1=\rho(u)\zeta(x)$, Eq.~(\ref{eq:Real}) is divided into the following two equations:  
\begin{align}
\label{Eq:zeta_5}
& \p_x^2\zeta-\xi^2\zeta-g \delta(x)\zeta=-\kappa^2\zeta \,, 
\\ 
\label{Eq:rho_5}
  & \rho'' - \frac{1+u^2}{u\, f}\, \rho'
  + \frac{1}{f}\, \left[ \frac{1}{u^2} + \frac{\hat{\mu}_0^2\, (1-u)}{4 u (1+u)}
    - \frac{\hat{\kappa}^2}{4 u}
    \right] \rho
   = 0~,
\end{align}
where $\kappa > 0$ is a separation constant,
and where $g:= \hat{g}L \,\pi^2\,T^2$, 
$\hmu_0 := \mu_0/ \pi T$, and $\hkappa := \kappa/ \pi T$ 
are the dimensionless quantities.

As $R_1$ depends on $x$, we can define the expansion parameter $\epsilon$ 
as the value of $|\Exp{{\cal O}}|$ at $x=l/2$, where $\Exp{{\cal O}}$ is 
the condensate of the dual field theory defined 
in the AdS/CFT duality. So, we can normalize 
\begin{align}
\label{normalization}
  & \lim_{u\to 0} \rho(u)/u = 1~,
& & \zeta(l/2) = 1~,
\end{align}
without loss of generality. 

To obtain an analytic solution of Eq.~(\ref{Eq:rho_5}), 
following the procedure of \cite{Herzog:2010}, we consider the following particular case:   
\begin{align}
\label{chemical_potential}
\hat{\kappa}^2\ll 1, \qquad 
\lim_{\hat{\kappa}\to 0}\hat{\mu}_0=2 \,.
\end{align}
Expanding Eq.~(\ref{Eq:rho_5}) as a series in 
$\hat{\kappa}^2$, we obtain $\rho$ as 
\begin{align}
\label{sol:rho_5}
& \rho=\frac{u}{1+u}-\hat{\kappa}^2\frac{u\ln(1+u)}{4(1+u)}+O(\hat{\kappa}^4), \nonumber \\
& \hat{\mu}_0=2+\frac{1}{2}\hat{\kappa}^2+O(\hat{\kappa}^4). 
\end{align}
Integrating Eq.~(\ref{Eq:zeta_5}) from $x=-\epsilon$ to $x=\epsilon$, we obtain 
\begin{align}
\label{bc:zeta}
\lim_{\epsilon\to 0}\{\p_x\zeta(+\epsilon)-\p_x\zeta(l-\epsilon)\}=g \zeta(0). 
\end{align}
Then, the general solution of Eq.~(\ref{Eq:zeta_5})~\footnote{In general, there is another solution such that $\zeta^2$ 
has a local minimum at $x=l/2$. However, this solution must oscillate at least once, so it takes larger value of $\kappa$ 
than the one of the solution~(\ref{sol:compactZeta}). 
This implies that it takes more energy than the solution~(\ref{sol:compactZeta}). } 
satisfying Eq.~(\ref{bc:zeta}) and the normalization (\ref{normalization})  
is written by   
\begin{align}
\label{sol:compactZeta}
& \zeta^2=\frac{1-\left({v_0}/{\kappa}\right)^2}{2}\cos{2\kappa(x-l/2)}+\frac{1+\left({v_0}/{\kappa}\right)^2}{2} 
\end{align}
with 
\begin{align}
\label{constant_g}
g =\frac{2(1-\left({v_0}/{\kappa}\right)^2)\kappa \sin\kappa l} 
{1+\left({v_0}/{\kappa}\right)^2+(1-\left({v_0}/{\kappa}\right)^2)\cos \kappa l}. 
\end{align}
\begin{figure}[ht]
  \centering
\includegraphics[bb=0 0 402 338, width=0.5\textwidth,clip]{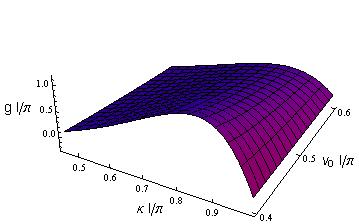}
 \caption{(color online). 
The plot of $g\, l$ as a function of $\kappa\, l$ and $v_0\, l$.
When $v_0$ is constant, there are two solutions for $\kappa$
as far as $g$ is less than a critical value $g_c$.
}
\label{fig:1}
\end{figure}
For given $v_0\, l$ and $g\, l$, we can find the parameters
$\kappa\, l$ that satisfy Eq.~(\ref{constant_g}). As shown in Fig.~\ref{fig:1}, there are two parameters 
$\kappa_<$, $\kappa_>~(\kappa_< < \kappa_>)$
for a given $v_0\, l$ when $g\, l$ is smaller than a critical value $g_c \,l$. 
The two parameters, however, merge at $g_c \,l$ and then beyond it, 
there is no solution satisfying Eq.~(\ref{constant_g}). 
The structure of the solutions is very similar to the one found in \cite{Hakim1997}, 
implying that the solution with $\kappa_>$ is unstable according to a saddle-node bifurcation theory.

\section{Higher order solutions and the free energy}
\label{sec:IV}
In this section we construct higher order solutions of Eqs.~(\ref{expansion}) 
and show that the solution with larger $\kappa$ is unstable 
by calculating the free energy. 
\subsection{The construction of $O(\epsilon)$ solutions}
At $O(\epsilon)$, the equations of motion are derived from 
Maxwell Eqs.~(\ref{eq:Maxwell}). 
From the condition~(\ref{asymt_M}) and the expansion~(\ref{expansion}), 
it follows that the r.~h.~s. of the $x$-component 
of Eqs.~(\ref{eq:Maxwell}) is $x$-independent at $O(\epsilon)$. 
This implies that $M_x^{(1)}$ is also $x$-independent. Making the ansatz $M_x^{(1)}=M_x^{(1)}(u)$ and $M_u^{(1)}=0$, Eqs.~(\ref{eq:Maxwell}) are reduced to 
\begin{align}
\label{eq:Mx_2_5}
(1-u^2){M_x''^{(1)}}-2u{M_x'^{(1)}}=\frac{v_0\rho^2}{2u^2} \,. 
\end{align}

From the boundary condition~(\ref{asymt_M}) and $\xi(l/2)=v_0$, 
the asymptotic boundary condition for $M_x^{(1)}$ should be $M_x^{(1)}(0)=0$. 
Imposing the regularity condition at the horizon $u=1$, 
we obtain the solution, up to $O(\hat{\kappa}^2)$, 
\begin{align}
\label{sol:Mx_2_5}
& M_x^{(1)}=\frac{v_0}{32(1+u)}\Biggl[8(1-\hat{\kappa}^2)+\hat{\kappa}^2\{(2\ln 2)(1+u)\ln(1-u) \nonumber \\
&-\ln(1+u)(4+2(1+u)\ln(1-u)-(1+u)\ln(1+u))\} \nonumber \\
&-2\hat{\kappa}^2(1+u)\mbox{Li}_2\left(\frac{1+u}{2}  \right)  
\Biggr]-\frac{v_0}{192}\{48-\hat{\kappa}^2(48+\pi^2-6(\ln 2)^2\} \nonumber \\
&+O(\hat{\kappa}^4) \,, 
\end{align}
where Li$_2$ is the Polygamma function~\cite{handbook}. The superfluid current $\Exp{J_x}$ is read off from 
the derivative of $M_x^{(1)}$ as 
\begin{align}
\label{R-current}
\Exp{J_x}\sim M_x'^{(1)}(0)=-\frac{v_0}{8}(2-\hat{\kappa}^2(1-\ln 2)) \,.  
\end{align}

The equation of motion for $M_t^{(1)}$ is given by 
\begin{align}
\label{eq:Mt_2_5}
& 4u\,\p_u^2M_t^{(1)}+\frac{\p_x^2M_t^{(1)}}{1-u^2} 
 =\frac{2\mu_0}{u(1+u)}\rho^2\zeta^2 
\nonumber \\
&=\frac{\mu_0\,\rho^2}{u(1+u)}\{(1-\nu^2)\cos{2\kappa(x-l/2)}+1+\nu^2\} \,,  
\end{align}
where here and hereafter we set $\nu =v_0/\kappa$ for brevity. 
Making the ansatz, $M_t^{(1)}(u,x)=\eta_{t0}(u)+\eta_{t1}(u)\cos{2\kappa(x-l/2)}$ and imposing the asymptotic boundary 
condition $\eta_{t1}(0)=0$ by Eq.~(\ref{asymt_M}), and the regularity conditions at the horizon, $\eta_{t0}(1)=\eta_{t1}(1)=0$, 
we obtain 
\begin{align} 
\label{sol:eta0}
& \frac{\eta_{t0}}{\pi T}=- C (1-u)+\frac{(1+\nu^2)(1-u)}{8(1+u)} \nonumber \\
&\,\,\,+\frac{\hat{\kappa}^2(1+ \nu^2)\{u-1+(1+u)\ln 2-2\ln(1+u)  \}}{16(1+u)}+O(\hat{\kappa}^4) \,, 
\end{align}
\begin{align}
\label{sol:eta1} 
& \frac{\eta_{t1}}{\pi T}=-\frac{(1-\nu^2)u(1-u)}{8(1+u)} \nonumber \\ 
&+\frac{\hat{\kappa}^2(1- \nu^2)u}{16(1+u)}\{1-u-(1+u)\ln 2+2\ln(1+u)   \}+O(\hat{\kappa}^4) \,,   
\end{align}
where $C$ is an integration constant determined later. 

\subsection{The free energy}
To regulate the action~(\ref{action_probe}), we need a counter term $S_{\text{ct}}$ defined by 
\begin{align}
\label{counter_term}
  & S_{\text{ct}}
  := \int d^4 x~\sqrt{-h}\, \left\{ \frac{2\, |\psi|^2}{L}
  + n^\mu \, \nabla_\mu |\psi|^2 
  \right\} \,, 
\end{align}
where $n^\mu$ is defined first as a unit outward normal vector 
to a $u= const.$ hypersurface with the induced metric $h_{ab}$, 
and where the limit $u \rightarrow 0$ is taken~\cite{Herzog:2010}. 
Evaluating the action~(\ref{action_probe}) on the 
onshell condition by using the equation of motion~(\ref{eq:scalar}) 
and using the asymptotic form of $\psi$~(\ref{asym_scalar}), 
we find 
\begin{align}
\label{onshell}
& S_{\text{os}}
= L^3(\pi T)^4 \int d^4x ( - 2|\beta|^2\ln u+\alpha \beta^\ast+\alpha^\ast \beta) \nonumber \\
&-\frac{L^2}{4}\int d^5x \sqrt{-g} F^2. 
\end{align}
Since we impose $\beta=0$ for the asymptotic boundary condition of $\psi$,  
the first term disappears, and there is no contribution from the scalar field.   

Using the Maxwell equation $\nabla_\nu F_{(0)}^{\nu\mu}=0$, we find  
\begin{align}
\label{expansion_F^2}
   S_{\text{os}}
  &= -L^2\int d^4 x \sqrt{-h}n^\mu F^{(0)}_{\mu\nu}
                                 ( M_{(0)}^\nu/2 + \epsilon M_{(1)}^\nu+\cdots)
\nonumber \\ 
  &- \frac{\epsilon^2 L^2}{2}\int d^4 x \sqrt{-h}n^\mu F^{(1)}_{\mu\nu}M_{(1)}^\nu
\nonumber \\
  &+ \frac{\epsilon^2 L^2}{2}\int d^5 x \sqrt{-g} M_{(1)}^\nu
  \nabla^\mu F^{(1)}_{\mu\nu}+\cdots~.
\end{align}
As shown in \cite{MNO2010}, the last term becomes zero under the asymptotic boundary condition $\beta=0$. 
Then, the free energy $\Omega=-S_{\text{os}}/\int dt$ becomes  
\begin{align}
\label{free_energy}
& \Omega=2\pi^2L^3T^2\int d^3 x M'^{(0)}_tM_t-\pi^2L^3T^2\int d^3 x M'^{(0)}_tM^{(0)}_t \nonumber \\
&+ \epsilon^2 \pi^2L^3T^2\int d^3 x M'^{(1)}_tM^{(1)}_t+\cdots. 
\end{align}
Here, we have used the fact that $M'^{(0)}_x=0$ and $M^{(1)}_x(u=0)=0$. 

In the limit $\hat{\kappa}\to 0~(T\to \infty)$, $M^{(0)}$ is independent of the scalar field configuration of $\psi$. Furthermore, 
our asymptotic condition $\mu=M_t(0, x)$ given by Eq.~(\ref{asymt_M}) implies 
that the difference of the free energy between the two 
solutions with $\kappa_<$ and $\kappa_>$ appears at the last term in Eq.~(\ref{free_energy}), up to $O(\epsilon^2)$.  
Substitution of the solution (\ref{sol:eta0}) and  (\ref{sol:eta1}) into Eq.~(\ref{free_energy}) yields 
\begin{align}
\label{free_energy1}
& \Omega=- \epsilon^2 \pi^4L^3T^4l\int dydz\times \Gamma, \nonumber \\ 
& \Gamma:=\left(C-\frac{1+ \nu^2}{8} \right)
\left(C - \frac{1+ \nu^2}{4}-\frac{(1- \nu^2)\sin \kappa l}{8\kappa l}\right).  
\end{align} 

The constant $C$ is determined by the ``orthogonality" 
condition~\footnote{For the derivation, see the Appendix A 
in Ref.~\cite{MNO2010}} 
\begin{align}
\int d^5 x \sqrt{-g} M_{(1)}^\nu\nabla^\mu F^{(1)}_{\mu\nu} =
\int d^5 x \sqrt{-g}M_{(1)}^\nu M^{(0)}_\nu R_1^2=0, 
\end{align}
where we used the Maxwell equation~(\ref{eq:Maxwell}) in the last line.  
Substituting Eqs.~(\ref{sol:Mx_2_5}), (\ref{sol:eta0}), and  (\ref{sol:eta1}) into the condition, we find 
\begin{align}
\label{orthogonal_cond}
& \left[\frac{(1+\nu^2)\sin \kappa l}{2\kappa l}+\frac{1-\nu^2}{4}+\frac{(1-\nu^2)\sin 2\kappa l}{8\kappa l}   \right]z_1 
\nonumber \\
& +\left[\frac{(1- \nu^2)\sin \kappa l}{2\kappa l}+\frac{1}{2}(1+ \nu^2) \right]z_0+O(\hat{\kappa}^2)=0
\end{align}
with the coefficients $z_0$ and $z_1$ given by 
\begin{align}
\label{coefficient}
& z_0=\frac{1}{192}(5+5\nu^2-48 C )+O(\hat{\kappa}^2), \nonumber \\
& z_1=-\frac{1}{192}(1-\nu^2)+O(\hat{\kappa}^2). 
\end{align}
 
The chemical potential can be expanded as 
\begin{align}
\label{chemical_expansion}
& \hat{\mu}=\frac{M_t(0,x)}{\pi T}=2+\frac{\hat{\kappa}^2}{2}
                               +\epsilon \hat{\mu}_1 + \cdots,  \nonumber \\
& \hat{\mu}_1:=- C + \frac{1+\nu^2}{8}.   
\end{align}
As $\mu$ and $T$ are fixed for the two solutions $\kappa_a~(a=>,\,<)$, the expansion 
parameter $\epsilon$ depends on $\kappa_a$. 
We evaluate the free energy (\ref{free_energy1}) in the case of 
the high temperature limit $\hat{\kappa}\to 0$ and $\epsilon^{1/2} \ll 1$. 
In this case, substituting Eq.~(\ref{chemical_expansion}) into 
Eq.~(\ref{free_energy1}) and eliminating $\epsilon$, we find 
\begin{align}
\Omega\simeq -(\hat{\mu}-2)^2\pi^4L^3T^4l\int dydz \,
\frac{\Gamma}{\hat{\mu}_1^2} \,. 
\end{align} 
This implies that the difference in the free energy 
between the two solutions is determined by the coefficient 
$\Gamma/\hat{\mu}_1^2$. 
 
We numerically solve Eq.~(\ref{orthogonal_cond}) and find the coefficient $\Gamma/\hat{\mu}_1^2$ for several cases. 
It turns out that the smaller value $\kappa$ takes, the larger positive 
value the coefficient $\Gamma/\hat{\mu}_1^2$ becomes. 
This indicates that the free energy with the smaller $\kappa~(\kappa_<)$ is smaller than the one with the larger $\kappa~(\kappa_>)$, 
as expected.
In the $g\, l = 0.6 \pi$, $v_0\, l = 0.4 \pi$ case,
for example, $\kappa_<\, l = 0.633\, \pi$,
$\kappa_>\, l = 0.957\, \pi$, 
and $\Gamma/\hat{\mu}_1^2=6.854$, $\Gamma/\hat{\mu}_1^2=5.917$, respectively. 

\section{Impurity and the spectral function} 
\label{sec:V}

So far, we have discussed properties of our holographic superfluid solutions  
with taking into account the effects of the background fluid flow $v_0$. 
In this section, we turn our attention to the effects of the impurity 
introduced by the repulsive potential. 

According to the dictionary of the AdS/CFT duality, the spectral function is 
derived from linear perturbation of the background solution obtained 
in Sec. \ref{sec:III} and \ref{sec:IV}.  
In general, the perturbed variables $\delta\psi$, $\delta A_\mu$ of the scalar field and the gauge field are coupled to each other. 
However, since the coupling constant is proportional to $\epsilon^{1/2}$~(see, for example, \cite{MaedaOkamura2008}), 
by taking the limit $\epsilon\to 0$, 
one can decouple those perturbation variables. 
Furthermore, by doing so, one can neglect the effects of the background flow 
$v_0$ on perturbations, and thereby manifest the effects of the impurity 
on the spectral function.  
In what follows, we consider, in this limit $\epsilon\to 0$, 
the fluctuations of the scalar field with the temperature $T$ and 
the coupling constant $g$ fixed, while the chemical potential $\mu$ taking various different values. 

Under the fixed gauge potential $A_\mu=\mu(1-u)\delta_{t\mu}$, 
the perturbed equation is simply given by Eq.~(\ref{eq:scalar}).
Here, note that the fluctuations of the superfluid velocity is 
encoded in the phase $\varphi$ of  the perturbed scalar field
$\delta\psi$~\footnote{The velocity corresponds to the derivative 
of $\varphi$ with respect to $x$. }.

Our background solution is independent of $t$ and homogeneous along 
the direction of $\vecY = (y, z)$, while it is inhomogeneous along 
the $x$-direction. In this case, the linear perturbation of 
the source term 
$\delta J = e^{- i \omega t + i \vecq \cdot \vecY} \delta J(\omega, \vecq\,|\, x) $ 
and the linear response of the condensates 
$\delta\Exp{\calO} = e^{- i \omega t + i \vecq \cdot \vecY} 
\delta \Exp{\calO(\omega, \vecq\,|\, x)}$ are related to each other, 
via the retarded response function $\retG(\omega, \vecq\,|\, x, x')$, 
as follows: 
\begin{align}
  & \delta \Exp{ \calO(\omega, \vecq\,|\, x) }
  = - \int dx'~\retG(\omega, \vecq\,|\, x, x')~
    \delta J(\omega, \vecq\,|\, x')
  ~.
\label{eq:def-response_func-Foourier}
\end{align}
Since the spectral function
\begin{align}
  & \rho(\omega, \vecq\,|\, x, x' )
  := - \Im\big[~\retG(\omega, \vecq\,|\, x, x' )~\big]~,
\label{eq:def-spectral_func}
\end{align}
specifies $\retG$ via its spectral representation, let us study the behavior of the spectral function $\rho$. 
In order to see the effects of the impurity, we focus on 
the linear perturbations $\vecq = \Vec{0}$ which are homogeneous 
along $\vecY = (y, z)$. (Hereafter we shall omit the argument 
$\vecq$, for simplicity.)

In the evaluation of $\retG(\omega\,|\, x, x' )$, it is convenient to 
introduce an appropriate complete orthonormal system (CONS) $\{ \chi_j \}$ 
in the $L^2(\mathbb{R})$ space along the $x$ direction. 
Then, Eq.~(\ref{eq:def-response_func-Foourier}) is expressed as 
\begin{align}
  & \delta \Exp{\calO_{i}(\omega)}
  = - \sum_{j}\, \retG_{\,ij}(\omega)~\delta J_{j}(\omega)
  ~, 
\label{eq:response-decomp-x}
\end{align}
where 
\begin{subequations}
\label{eq:response-decomp-x-all}
\begin{align}
  & \delta J_{j}(\omega)
  := \int dx~\chi_j^*(x)~\delta J(\omega\,|\, x)~,
\label{eq:delJ-decomp-x} \\
  & \retG_{\,ij}(\omega)
  := \int dx\, dx'~\chi_i^*(x)~
    \retG(\omega\,|\, x, x')~\chi_{j}(x')
  ~,
\label{eq:retG-decomp-x} \\
  & \retG(\omega\,|\, x, x' )
  = \sum_{i,j}\, \chi_i(x)~\retG_{\,ij}(\omega)~\chi_{j}^*(x')
  ~.
\label{eq:retG-decomp-x-rev}
\end{align}
\end{subequations}

Now let us derive Eq.~(\ref{eq:response-decomp-x}) by using 
{\it holographic method}. We seek for a solution of the perturbed 
scalar field $\delta\psi$ in the form of separation of variables 
$\delta\psi = e^{-i \omega t}\, U(u)\, X(x)$. The equation of motion for 
$\delta \psi$
becomes the following set of equations:
\begin{subequations}
\label{eq:EOM-linear_res_field-all}
\begin{align}
  & 
  \left[ - \frac{d^2}{dx^2} + g \sum_{n=-\infty}^\infty \delta(x-nl)\,
  \right] X = \kappa^2\, X  ~,
\label{Eq:pert1} \\ 
  & \left[\, \frac{d}{du} \frac{f(u)}{u} \frac{d}{du}
  + \frac{ \left\{ \hat{\omega} + \hat{\mu} (1-u) \right\}^2 }{4 u^2 f(u)}
      + \frac{4 - \hat{\kappa}^2 u}{4 u^3}\, \right] U = 0~,
\label{Eq:pert2} 
\end{align}
\end{subequations}
where $\homega := \omega/\pi T$, $\hmu := \mu/\pi T$, and $\kappa > 0$.

In the limit $\epsilon\to 0$, the background solution in previous sections 
affects Eq.~(\ref{Eq:pert2}) only through $\hat{\mu}$. 
Note that Eq.~(\ref{Eq:pert2}) is equivalent to Eq.~(\ref{Eq:rho_5})
when $\Hat{\omega}=0$.

Let us first analyze Eq.~(\ref{Eq:pert1}). 
Since Eq.~(\ref{Eq:pert1}) corresponds to an energy eigenvalue problem 
of the Kronig-Penney model consisting of $\delta$-function barriers 
with certain period, it should admit, as a solution, Bloch states 
\begin{align}
  & X_k(x + l) = e^{i k l}\, X_k(x)
& & \left( - \infty < k < \infty \right)~. 
\label{eq:Bloch_cond}
\end{align}
Demanding the orthonormal condition, 
\begin{align}
  & \int^\infty_{-\infty} dx~X_k^*(x)\, X_{k'}(x)
  = \delta(k - k'\,)
  ~,
\label{eq:def-AdS_CFT-x-orthonormal}
\end{align}
we find the solutions to Eqs.~(\ref{Eq:pert1}) and (\ref{eq:Bloch_cond}) as  
\begin{align}
  & X_k(0 < x < l)
  = \left\{ 8 \pi \, \sin \kappa\, l~\sin k\, l~(dk/d\kappa) \right\}^{-1/2}
\nonumber \\
  &\hspace*{0.5truecm} \times
  \big\{
    \left( e^{i k l} - e^{- i \kappa l} \right)\, e^{i \kappa x}
    - \left( e^{i k l} - e^{i \kappa l} \right)\, e^{- i \kappa x}
  \big\}
  ~.
\label{eq:Kronig_Penney-sol-X}
\end{align}
Here $\kappa$ is a positive definite function $\kappa(k) > 0$ 
of the wave number $k$ and is determined by~\cite{Kittel},  
\begin{align}
   - 1 \le  \cos k \ell
  &= \cos \kappa \ell
  + \frac{g\, \ell}{2}~\frac{ \sin \kappa \ell }{\kappa \ell}
  \le 1
  ~.
\label{eq:KP-band}
\end{align}
Note that $\kappa(k) = \kappa(-k)$, $X_k^*(x) = X_{-k}(x)$. 

\begin{figure}[ht]
\centering
\includegraphics[bb=0 0 360 239, width=0.5\textwidth,clip]{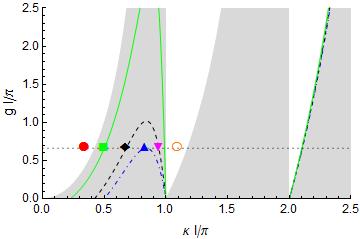}
\caption{(color online). The band structure given by the repulsive potential
(\ref{external_potential}).
The shaded region is the allowed region in which 
Eq.~(\ref{eq:KP-band}) holds, and the blank region is forbidden one.
The dot-dashed~(blue) curve represents the function $g$ given by Eq.~(\ref{constant_g})
with $v_0\, l = \pi/2$, while the dashed~(black) and solid~(green) curves 
represent the ones with  $v_0\, l = 0.421\,\pi$ and  $v_0\, l =0.225\,\pi$, respectively.
The horizontal black dotted line expresses the local maximum value 
($g\, l = 0.663\, \pi$) of the blue dot-dashed curve. 
The six points 
on the horizontal dotted line correspond to 
$\kappa_* l/\pi := T l\, \hkappa_*(\hmu)$ for the value of 
the chemical potential $\mu l/\pi$~respectively,
$2.060$~(\textcolor{red}{\Large$\bullet$}, red),
$2.120$~(\textcolor{green}{$\blacksquare$}, green), 
$2.215$~(\textcolor{black}{$\blacklozenge$}, black),
$2.321$~(\textcolor{blue}{\large$\blacktriangle$}, blue), 
$2.400$~(\textcolor{magenta}{\large$\blacktriangledown$}, magenta), 
$2.519$~(\textcolor{orange}{\Large$\circ$}, orange)
when $Tl= 1$ and $g\, l = 0.663\, \pi$. 
}
\label{fig:band}
\end{figure}

The condition (\ref{eq:KP-band}) places a restriction on the range of $\kappa$ for a given $g$,
yielding a band structure shown in Fig.~\ref{fig:band}. 
The shaded region represents the allowed region satisfying Eq.~(\ref{eq:KP-band}), while the blank region corresponds to the forbidden region.
Let us denote by $\kappa_n$ the bottom of the $n$-th allowed band.  
Then, the first and the second allowed bands are given respectively by 
$\kappa_1 < \kappa < \pi/l$ and $\kappa_2 < \kappa < 2 \pi/l$. 
For $g\, l = 0.663\, \pi$, $\kappa_1\, l = 0.423\, 
\pi$, $\kappa_2\, l = 1.175\, \pi$.

In the case of $v_0\, l = \pi/2$, 
the dot-dashed (blue) curve
represents the function $g$ of $\kappa$ 
given by Eq.~(\ref{constant_g}), while the dashed~(black) and solid~(green) curves 
represent the ones with  $v_0\, l = 0.421\,\pi$ and  $v_0\, l =0.225\,\pi$, respectively.

Let us turn to Eq.~(\ref{Eq:pert2}). Hereafter, by $U_{k}$ we denote $U$ that couples to $X_{k}$\footnote{
Note that in what follows, we assume that the orthonormal Bloch states 
(\ref{eq:Kronig_Penney-sol-X}) form a CONS on the $L^2(\mathbb{R})$ space 
defined by  
\begin{align}
  & \int^\infty_{-\infty} dk~X_k(x)\, X_k^*(x')
  = \delta(x - x'\,)
  ~.
\label{eq:def-AdS_CFT-x-complete}
\end{align}
} 
so that $\delta\psi \propto X_{k}\, U_{k}$.  
Near the AdS boundary $u \sim 0$, it becomes
\begin{align}
  & \left( \frac{d}{du}\, \frac{1}{u}\, \frac{d}{du}
  + \frac{1}{u^3} \right) U_{k} \sim 0~,
\end{align}
and therefore its solution behaves asymptotically as 
\begin{align}
  & U_{k}(u)
  \sim - \alpha(\hkappa(k), \hmu, \homega)\, u 
       + \beta(\hkappa(k), \hmu, \homega)\, u\ln u~.
\end{align} 
Here we have taken into consideration the fact that the wave number $k$ contributes to the above formula only through $\hkappa(k)$ in Eq.~(\ref{Eq:pert2}).

Note that $\alpha$ and $\beta$ are related in a certain manner 
so that at the horizon $u \rightarrow 1$, $U_{k}$ satisfies the in-going 
boundary conditions. 
This, together with the fact that Eq.~(\ref{Eq:pert2}) 
with different wave number $k$ is independent 
of each other, implies that the expectation value $\alpha(\hkappa(k))$ 
couples only to the source $\beta(\hkappa(k))$ with the same 
wave number $k$. 
Therefore we can diagonalize $\retG_{ij}(\omega)$ 
in Eq.~(\ref{eq:response-decomp-x}) so that $\retG_{k k'}(\homega) = \retG_{\hkappa}(\homega)\, \delta(k - k' )$, 
with $\retG_{\hkappa}(\homega)$ given below. Note that $\retG_{\hkappa}(\homega)$ 
is labeled only by $\hkappa$, since $\alpha$ 
and $\beta$ has a dependency on $k$ only through $\hkappa$. 
Thus, we obtain the retarded response function as 
\begin{subequations}
\label{eq:Green-all}
\begin{align}
  & \retG(\homega\,|\,x, x')
  = \int^\infty_{-\infty} dk~\retG_{\hkappa}(\homega)\,
  X_{k}(x)\, X^{\ast}_{k}(x')
\nonumber \\
  &\hspace*{0.5truecm}
  = 2 \int^\infty_{0} dk~\retG_{\hkappa}(\homega)\,
  \Re\left[\, X_{k}(x)\, X^{\ast}_{k}(x')\, \right]
  ~,
\label{Green} \\
  & \retG_{\hkappa}(\homega)
  := - \frac{ \alpha(\hkappa(k), \hmu, \homega) }
          { \beta(\hkappa(k), \hmu, \homega) }~.  
\label{eq:def-Green_diagonal}
\end{align}
\end{subequations}

The behavior of the retarded response function $\retG_{\hkappa}(t)$ 
with real time $t$ is determined by the singularity structure 
of $\retG_{\hkappa}(\homega)$ with respect to the complex $\homega$.  
We can find the relaxation time scale of the condensate 
by inspecting quasinormal (QN) frequency 
$\homega = \hOmega_{\tQNM}(\hkappa, \hmu)$, which is a solution 
to $\beta(\hkappa, \hmu, \homega ) = 0$ and provides 
the poles of $\retG_{\hkappa}(\homega)$. 
Furthermore, we find a critical point from either the highest value of 
$\hmu$ ($\hmu = \hmu_\ast(\hkappa)$) or the lowest value of 
$\hkappa$ ($\hkappa = \hkappa_\ast(\hmu)$) that solves the source free 
condition for linear perturbation (response) field in the stationary case. 
As shown later, the spectral function crucially depends on 
the background parameter $\hat{\kappa}_\ast(\Hat{\mu})$.

Now let us see the behavior of the spectral function 
at the point $x = x' = l/2$, avoiding in particular the location of 
the impurity itself. 
Define $\rho(\Hat{\omega}) := \rho(\Hat{\omega}\,|\, l/2, l/2)$. Then 
from Eqs.~(\ref{eq:def-spectral_func}), (\ref{eq:Kronig_Penney-sol-X}), 
and (\ref{Green}), we obtain: 
\begin{subequations}
\label{eq:def-spectral_func-mid_pt-all}
\begin{align}
  & \rho(\Hat{\omega})
  = - \int_{\band} \frac{d\kappa}{\pi}~\Im\left[\, \retG_{\hkappa}(\homega)
  \, \right]\, \calW(\kappa\, l,\, g\, l)
  ~,
\label{eq:def-spectral_func-mid_pt} \\
  & \calW(\sigma,\, g\, l)
  := \frac{ \sin\sigma + (g\, l/2\, \sigma)\, (1 - \cos\sigma) }
          {\sin k\, l}\,
\label{eq:def-spectral_func-weigh-I} \\
  &\hspace*{1.5truecm}
  = \sqrt{ \frac{ 2\, \sigma + g\, l\, \tan(\sigma/2) }
                { 2\, \sigma - g\, l\, \cot(\sigma/2) } }
  ~, 
\label{eq:def-spectral_func-weigh-II}
\end{align}
\end{subequations}
where the domain of integration $\band$ satisfies Eq.~(\ref{eq:KP-band}) 
with $\kappa > 0$ and is determined by $g\, l$.  
We list below the basic properties of the spectral function 
$\rho(\homega)\, l$, obtained by inspecting 
Eq.~(\ref{eq:def-spectral_func-mid_pt-all}): 
\begin{itemize}
\item The spectral function $\rho(\homega)\, l$ is specified by 
  $\hmu$, $T\, l$, and $g\, l$. 
  (Note that $\hkappa = \kappa\, l/(\pi\, T\, l)$).
\item The dependency on $\hmu$ and $T\, l$ is determined by  
  {$\retG_{\hkappa}(\homega)$}. 
  Eq.~(\ref{Eq:pert2}), which provides $\retG_{\hkappa}(\homega)$, is 
  the same as the equation for linear perturbations with wave number $k$ 
  on the homogeneous background field with no impurity. 
  (To see this, replace $\hkappa$ with $k/\pi T$.)
  From the dependency of $\retG_{\hkappa}(\homega)$ on low-$\homega$ and 
  $\hkappa$ around the critical point $\hmu \lesssim \hmu_*(\hkappa)$, 
  it turns out that our holographic superconductor exhibits the critical 
  dynamics of model A \cite{Maeda:2009wv}. 
 (See Ref~\cite{hohenberg_halperin} for a review of dynamic critical phenomena, and also \cite{DC_textbook} for the introduction of dynamic critical phenomena. 
  The study of dynamic critical phenomena in the AdS/CFT context is given 
  by, e.g., \cite{Maeda:2008hn,Buchel:2010gd,Natsuume:2010bs}.) 
\item The effects of the impurity appear via the band structure $\band$ and 
  the weight $\calW(\kappa\, l,\, g\, l)$. 

\item Due to ``$\sin k\, l$'' in the denominator of 
Eq.~(\ref{eq:def-spectral_func-weigh-I}) (or Eq.~(\ref{eq:def-spectral_func-weigh-II})), the weight $\calW$ diverges at the edge of the band, where  
$\cos k\, l = \pm 1$. 
Since $1/\sin k\, l \propto dk/d\kappa$ from Eq.~(\ref{eq:KP-band}), 
the divergence of $\calW$ is related to that of the state density 
per unit ``energy'' $E := \kappa^2$, i.e., $dk/dE$.  
This is a reminiscent of the fact that in superfluid, local dynamical 
response function exhibits a singular behavior around the critical 
velocity\cite{KatoWatabe1,KatoWatabe2}. 
The divergence of state density at the critical velocity is the origin of 
the singular behavior. 

\end{itemize}

\begin{figure}[ht]
\centering
\includegraphics[bb=0 0 1187 675, width=0.5\textwidth,clip]{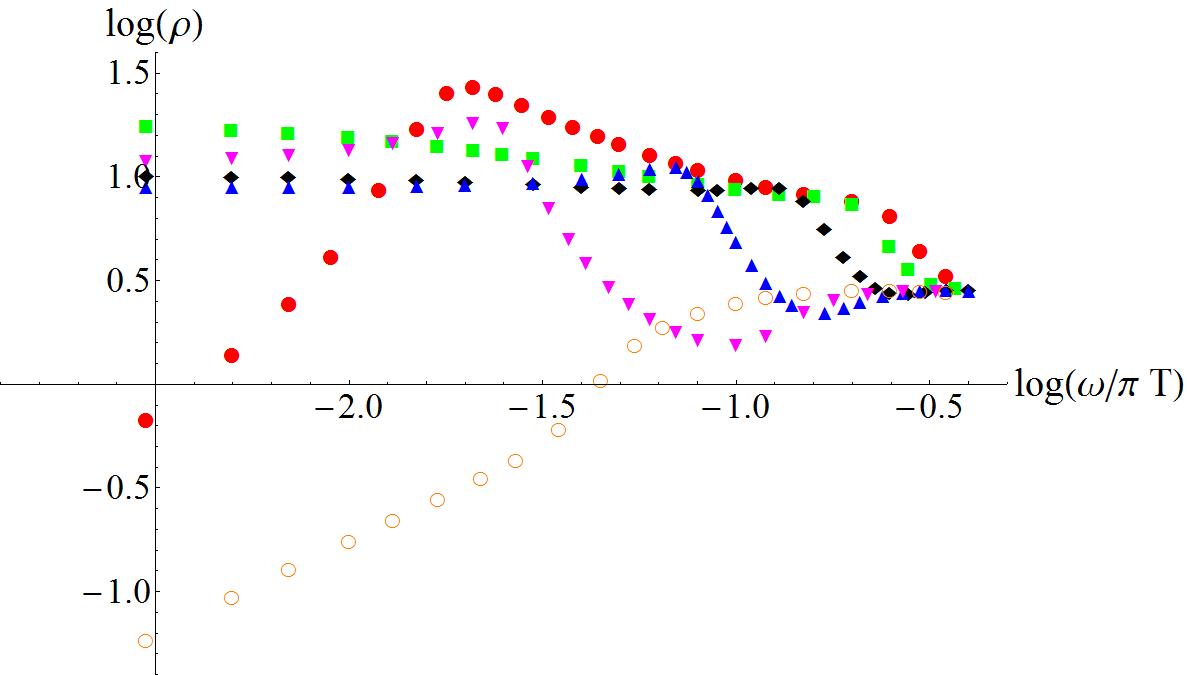}
\caption{(color online). The Log-Log plot of the spectral function $\rho(\homega)$ 
is shown for various chemical potentials 
$\mu\, l/\pi$~($2.060$~(\textcolor{red}{\Large$\bullet$}, red),
$2.120$~(\textcolor{green}{$\blacksquare$}, green), 
$2.215$~(\textcolor{black}{$\blacklozenge$}, black),
$2.321$~(\textcolor{blue}{\large$\blacktriangle$}, blue), 
$2.400$~(\textcolor{magenta}{\large$\blacktriangledown$}, magenta), 
$2.519$~(\textcolor{orange}{\Large$\circ$}, orange)
when $Tl = 1$ and $g\, l = 0.663\, \pi$. 
Note that these points with the colors (red, green, black, blue, magenta, and orange) correspond respectively to the points 
on the horizontal black dotted line with the same colors in Fig.~\ref{fig:band}}.   
\label{fig:spectral_function_1}
\end{figure}

In Fig.~\ref{fig:spectral_function_1},
we plot the spectral function at {$x = l/2$}, $\rho(\homega)$, 
for various chemical potentials $\mu$
for {$T\, l = 1$ and $g\, l = 0.663\, \pi$ case.} 
Then, the behavior of the spectral function is qualitatively determined
by the position of $\kappa_\ast := \pi T\, \hkappa_*$.
When $\kappa_\ast\, l/\pi = (T\, l)\, \hkappa_\ast$ 
is in the blank region in Fig.~\ref{fig:band}, 
the spectral function rapidly increases as $\homega$ 
increases~(see the red~($\kappa_\ast\, l = 0.35\, \pi$)
and orange curve~({$\kappa_\ast\, l = 1.1\, \pi$}) in Fig.~\ref{fig:spectral_function_1}).
On the other hand, when {$\kappa_\ast\, l/\pi$}
is in the gray region in Fig.~\ref{fig:band},
the spectral function does not change rapidly in the low frequency region,
and it decays around ${\homega} \sim 0.1$~
(see the other curves in Fig.~\ref{fig:spectral_function_1}).
\begin{figure}[ht]
\centering
\includegraphics[bb=0 0 743 437, width=0.5\textwidth,clip]%
{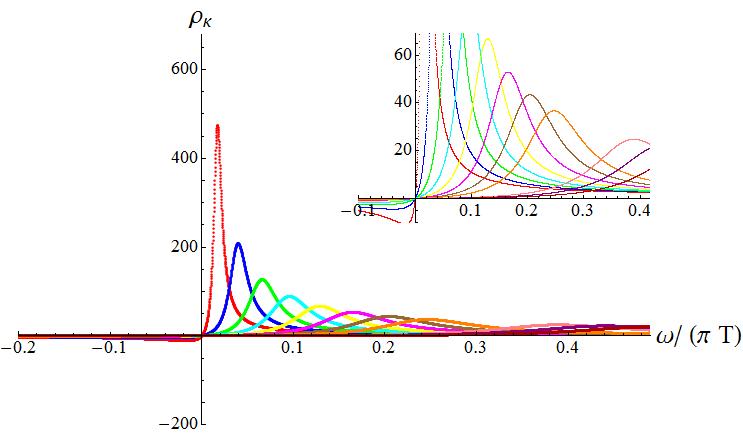}
\caption{(color online). 
Plots of $\rho_{\hkappa}(\homega)$ for various different values of $\hkappa$
with $\hkappa_* = 0.35$, corresponding to the red plot (\textcolor{red}{\Large$\bullet$})
in Fig.\ref{fig:spectral_function_1} The location of peak shifts to the right, 
as $\hkappa$ increases from $\hkappa = 0.423$ with the difference 
$\Delta \hkappa = 0.08$ between every adjacent two peaks. 
When $T\, l = 1$, 
$\hkappa \simeq 0.423$, which gives the peak closest to the origin, 
corresponds to, $\kappa_1 l/\pi \sim 0.423$, the bottom of the first 
allowed band for $g\, l = 0.663\, \pi$. 
One can find the disappearance of peaks around $\homega = 0.32$ due to 
the band gap for $\hkappa = 1.063$ and $1.143$. 
 }
\label{fig:spectral-each_k-035}
\end{figure}
\begin{figure}
\centering
\includegraphics[bb=0 0 633 379, width=0.5\textwidth,clip]%
{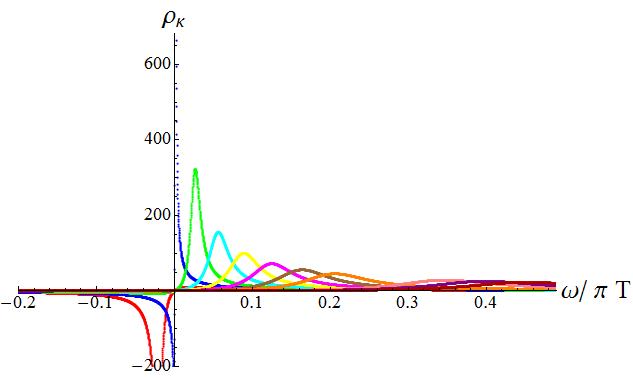}
\caption{(color online). Plots of $\rho_{\hkappa}(\homega)$ for various different values of $\hkappa$ 
with $\hkappa_* = 0.50$, corresponding to the green plot (\textcolor{green}{$\blacksquare$})
in Fig.\ref{fig:spectral_function_1}.  
The plots with the same color in this figure and 
Fig.~\ref{fig:spectral-each_k-035} are for the same values of 
$\hkappa$.  
Due to the large value of $\hkappa_*$ (or $\hmu$) than that for 
Fig.~\ref{fig:spectral-each_k-035}, every peak is shifted to 
the left compared to the location of the corresponding peak 
in Fig.~\ref{fig:spectral-each_k-035}.  
The disappearance of peaks due to the band gap is 
also shifted to around $\homega = 0.29$. 
Note also that there is a mode that has a peak arbitrarily close 
to $\omega =0$, as $0.423 < \hkappa_*$.  
}
\label{fig:spectral-each_k-050}
\end{figure}
\begin{figure}[ht]
\centering
\includegraphics[bb=0 0 633 379, width=0.5\textwidth,clip]%
{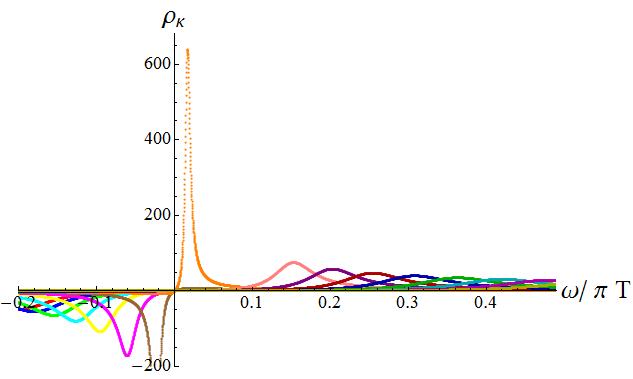}
\caption{(color online). 
Plots of $\rho_{\hkappa}(\homega)$ for various different values of $\hkappa$ 
with {$\hkappa_* = 0.95$}, corresponding to the magenta plot (\textcolor{magenta}{\large$\blacktriangledown$})
in Fig.\ref{fig:spectral_function_1}. The location of every peak is shifted further to the 
left, and the disappearance of peaks due to the band gap is 
now around $\homega = 0.06$, which forms the slope around 
$\log (\omega/\pi T) = - 1.25$ in Fig.\ref{fig:spectral_function_1}.
}
\label{fig:spectral-each_k-090}
\end{figure}
\begin{figure}
\centering
\includegraphics[bb=0 0 633 379, width=0.5\textwidth,clip]%
{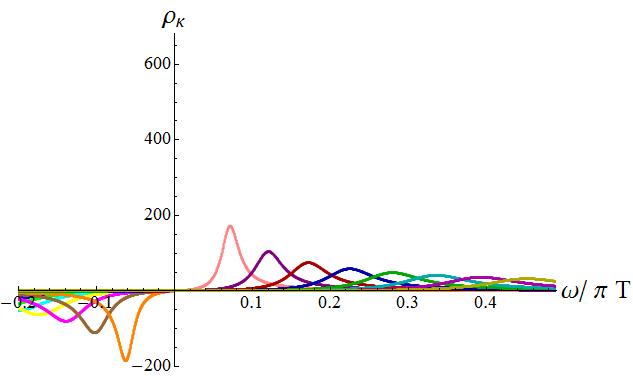}
\caption{(color online). 
Plots of $\rho_{\hkappa}(\homega)$ for various different values of $\hkappa$ 
with $\hkappa_* = 1.10$, corresponding to the magenta plot (\textcolor{orange}{\Large$\circ$})
in Fig.\ref{fig:spectral_function_1}.  
The disappearance of peaks due to the band gap is 
now around $\homega \sim 0$, and 
$\rho(\homega \sim 0^+)$ takes vanishingly small values, corresponding 
to the behavior of the orange plot in Fig.\ref{fig:spectral_function_1}. 
}
\label{fig:spectral-each_k-110}
\end{figure}

The behavior of $\rho(\homega)$ shown in Fig.~\ref{fig:spectral_function_1} 
can be explained by inspecting the dependency of 
$\rho_{\hkappa}(\homega)
:= - \Im[\, \retG_{\hkappa}(\homega) \,]
$ on 
$\hmu$ (or $\hkappa_*$) and $\kappa$, and by taking the existence of 
a band gap into consideration.

We first note that $\retG_{\hkappa}(\homega)$ is expressed in terms of 
its QN frequencies $\{ \hOmega_{\tQNM,j} \}_{j=0, 1, 2, \cdots}$ as 
\begin{align*}
  & \retG_{\hkappa}(\homega)
  = \sum_j\, \frac{ a_j(\hkappa, \hmu) }
                  { \homega - \hOmega_{\tQNM,j}(\hkappa, \hmu)  }
  + \cdots~,
\end{align*}
where $a_j$ denotes the residues of $\retG_{\hkappa}(\homega)$ at 
$\homega = \hOmega_{\tQNM,j}$. 
Now, let $\Hat{\bmOmega}(\hkappa, \hmu)$ be the QN frequencies 
whose imaginary parts take the smallest in their absolute value. 
From the above expression of $\retG_{\hkappa}(\homega)$, we find that 
for real $\homega$, such an $\Hat{\bmOmega}(\hkappa, \hmu)$ can contribute 
most to $\retG_{\hkappa}(\homega)$, hence to $\rho_{\hkappa}(\homega)$, 
and we have 
\begin{align*}
  & \retG_{\hkappa}(\homega)
  \sim \frac{ \bma(\hkappa, \hmu) }
            { \homega - \Hat{\bmOmega}(\hkappa, \hmu) }
  = \frac{ \bma\, ( \homega - \Hat{\bmOmega}^* ) }
    { |\, \homega - \Hat{\bmOmega}\, |^2 }
  ~. 
\end{align*}
Then, the behavior of $\rho_{\hkappa}(\homega)$ can be understood 
in terms of $\Hat{\bmOmega}(\hkappa, \hmu)$ via the following formula: 
\begin{align}
  & \rho_{\hkappa}(\homega)
  \sim - \frac{ \Im(\bma)~\homega }
              { |\, \homega - \Hat{\bmOmega}(\hkappa, \hmu)\, |^2 }
  ~, 
\label{eq:rho_kappa-by_QNM-approx}
\end{align}
where we have set $\Im[\, \bma\, \Hat{\bmOmega}^*\, ] = 0$ as 
$\rho_{\hkappa}(\homega = 0) = 0$.

We find that $\Hat{\bmOmega}(\hkappa, \hmu)$ has the following properties: 
\begin{itemize}
\item When $\hkappa = \hkappa_*(\hmu)$, $\Hat{\bmOmega} = 0$\footnote{
This is indicated from the fact that $\hmu_*$ ($\hkappa_*$) satisfies 
the source free condition of Eq.~(\ref{Eq:pert2}) with $\homega = 0$ 
(i.e., boundary conditions at the infinity for QN frequencies).  
For $\hkappa = 0$, such a massless mode may be viewed as the emergence of 
a holographic Nambu-Goldstone mode in the condensation phase~\cite{Amado:2009ts}. 
}, 
and when $\hkappa \gtrless \hkappa_*(\hmu)$, 
$\Im[\, \Hat{\bmOmega}(\hkappa, \hmu)\, ] \lessgtr 0$. 
That is, perturbations with $\hkappa > \hkappa_*(\hmu)$ are stable ones, 
while those with  $\hkappa < \hkappa_*(\hmu)$ are unstable ones. 

\item Since $\hkappa_*(\hmu)$ is an increasing function, 
as $\hmu$ increases, the stable perturbations change to unstable ones, 
and at the marginal limit $\hmu \nearrow \hmu_*(\hkappa)$ 
($\hkappa_*(\hmu) \nearrow \hkappa$), $\Hat{\bmOmega} \to 0$~\cite{Maeda:2009wv,Amado:2009ts}. This implies that $\Hat{\bmOmega}$ is continuous with $\hmu$. 

\item For the stable perturbations with $\hkappa > \hkappa_*(\hmu)$, 
both the magnitude of the real part of $\Hat{\bmOmega}(\hkappa, \hmu)$ 
and that of the imaginary part are increasing functions with respect 
to $\hkappa$. 
\end{itemize}
From the observations above, we can expect $\rho_{\hkappa}(\homega)$ to 
behave as follows: 
\begin{itemize}
\item 
As Eq.~(\ref{eq:rho_kappa-by_QNM-approx}) shows, 
$\rho_{\hkappa}(\homega)$ possesses a Lorentzian peak of width 
about $|\Im[\, \Hat{\bmOmega}(\hkappa, \hmu)\, ]|$ 
at $\homega = \Re[\, \Hat{\bmOmega}(\hkappa, \hmu)\, ]$. 
(In what follows we assume $\Re( \Hat{\bmOmega} ) > 0$.)  
\item As the marginal limit $\hmu \nearrow \hmu_*(\hkappa)$ 
($\hkappa_*(\hmu) \nearrow \hkappa$) is approached, 
the peak of $\rho_{\hkappa}(\homega)$ is getting narrow in width and sharp,  
and its location is approaching $\homega = 0$.

\item The modes which become unstable beyond the marginal limit possess 
a negative peak in the range $\homega<0$. 
\end{itemize}
Figs.~\ref{fig:spectral-each_k-035}-\ref{fig:spectral-each_k-110} show 
that our expectations about the behavior of $\rho_{\hkappa}(\homega)$ listed 
above are in fact true.

As in Eq.~(\ref{eq:def-spectral_func-mid_pt-all}), the local spectral function 
$\rho(\homega)$ can be obtained by the (weighted with $\calW$) summation of 
$\rho_{\hkappa}(\homega)$.  
Then, from Figs.~\ref{fig:spectral-each_k-035}-\ref{fig:spectral-each_k-110}, 
we can find the behavior of $\rho(\homega)$ in 
Fig.~\ref{fig:spectral_function_1} as follows: 
\begin{itemize}
\item $(T\, l)\, \hkappa_*(\hmu) < \kappa_1\, l/\pi$~
  (Fig.~\ref{fig:spectral-each_k-035})
 
  The minimum value of $\kappa$ is $\kappa_1$, whose peak location,  
  $\homega_1 := \Re[\, \Hat{\bmOmega}(\kappa_1/\pi T, \hmu)\, ]$, 
  is the closest to $\homega=0$ and the tallest among others.  
  As $\hkappa$ increases, the peak is shifted to the right and 
  its shape becomes short in height and wide width. Then, 
  $\rho(\homega)$ obtained by summing up such short and wide profiles 
  increases with $\homega$ in a neighborhood $\homega \gtrsim 0$, 
  admits a peak around $\homega_1$, and then monotonically decreases 
  as the red plots in Fig.\ref{fig:spectral_function_1}. 
\item $\kappa_{1}\, l/\pi < (T\, l)\, \hkappa_*(\hmu) < 1$~
  (Figs.~\ref{fig:spectral-each_k-050}, \ref{fig:spectral-each_k-090})

  In this case $\homega_{1} < 0$. Since $\hkappa_*(\hmu)$ is 
  in the first allowed band, there exists a mode of $\hkappa$ which can 
  arbitrarily be close to $\hkappa_*$, and in the range $\homega > 0$, 
  sharp peaks continue to $\homega = 0$. For this reason, $\rho(\homega)$ 
  does not show an increasing behavior in $\homega$ in a neighborhood 
  of $\homega = 0$ shown in Fig.~\ref{fig:spectral_function_1}.

  Also in that neighborhood, the disappearance of peaks due to the band gap 
  makes $\rho(\homega)$ decreasing, reflected in the plots on the 
  decreasing slope (with colors, green, black, blue, and magenta) 
  in Fig.~\ref{fig:spectral_function_1}.  

\item $1 < (T\, l)\, \hkappa_*(\hmu) < \kappa_{2}\, l/\pi$~
  (Fig.~\ref{fig:spectral-each_k-110})

  Due to the band gap, there are no peaks in a neighborhood of $\homega = 0$;  
  no contribution to $\rho(\homega)$ in the neighborhood of $\homega = 0$.   
  Therefore the region of $\homega \gtrsim 0$ looks similar 
  to Fig.~\ref{fig:spectral-each_k-035}. As a result, the plot of 
  $\rho(\homega)$ also becomes similar to that of 
  $(T\, l)\, \hkappa_*(\hmu) < \kappa_{1}\, l/\pi$ 
  (see the red plots in Fig.~\ref{fig:spectral-each_k-035}).  

\end{itemize}

Although in this section we focused on the impurity and spectral function 
with neglecting the effects of the background flow $v_0$, 
in order to obtain some insights into critical behavior in our holographic 
superfluid models, we may give the following interpretation 
on the relation between $\kappa_*$ and the background flow $v_0$. 

By Eq.~(\ref{constant_g}),
$\kappa_\ast\, l~(= \pi\, T\, l\, \hat{\kappa}_\ast)$
is related to the background fluid flow velocity $v_0\, l$.
Then, the parameter values $\kappa_\ast\, l/\pi = 0.5$~(green),
$0.677$~(black), and $0.840$~(blue) 
in Fig.~\ref{fig:spectral_function_1} correspond respectively
to $v_0\, l/\pi = 0.225$, $0.421$, and $0.5$.  

As explained in Sec.~III, there are two solutions $\kappa_\ast$ satisfying Eq.~(\ref{constant_g}) when $v_0<v_c$. 
In the above case,
$v_c\, l \simeq 0.5\, \pi$,
so the blue curve in Fig.~\ref{fig:spectral_function_1} corresponds to the almost critical case 
where the two solutions merge. 

As $v_0$ approaches the threshold $v_c$ from below, a small hill accompanied 
with a steep slope emerges in the spectral function $\rho(\homega)$ around 
$\homega = \omega/\pi T \sim 0.1$~({blue} curve).
The divergent state density at the edge of the band gap, $\kappa\, l = \pi$ induces
the small hill, while the steep slope is caused by the band gap above, $\kappa\, l > \pi$, as discussed above.  
Fig.~\ref{fig:spectral_function_1} shows that the slope becomes steep as $v_0$ approaches the threshold 
$v_c$ from below. This reflects the fact that the imaginary part of the QN frequencies, 
$\Hat{\bmOmega}(\hkappa, \hmu)$ in the band gap, $\kappa l>\pi$ becomes small as 
$\kappa_\ast$ approaches the edge of the band gap, $\kappa l=\pi$.

\section{summary and discussions}
\label{sec:VI}
We have investigated holographic models of one-dimensional superfluid flow 
solutions in the presence of an external repulsive potential. 
Our solutions are a generalization of the solution 
of the GP equation~\cite{Hakim1997} to 
the strongly coupled case. Our solutions possess the properties 
very similar to those found in GP equation: (i) There are 
two solutions below the critical value $g_c$ of the coupling constant $g$, 
and they merge at $g_c$. (ii) The free energy of the 
solution with steeper configuration is higher than the other solution, 
implying that the solution with steeper configuration is unstable.  
 
We have also studied the spectral function derived from the perturbation of the steady superfluid flow solution. 
Due to the band structure generated by the periodic repulsive potential, 
the qualitative features of the spectral function are essentially determined 
by the parameter $\kappa_\ast$ which satisfies $\beta(\kappa_\ast, \omega=0)=0$. 
As shown in Fig.~\ref{fig:spectral_function_1}, as $\kappa_\ast$ approaches 
the edge of the band from the left hand side, a small 
hill appears. 
This reflects the fact that the state density, hence $\calW$, 
diverges at the edge.

For the solution of one-dimensional superfluid flow~\cite{Hakim1997}, the spectral function 
$\rho(\omega)$ of the local density fluctuation was derived by solving Bogoliubov equation of the GP equation~\cite{KatoWatabe1, KatoWatabe2}. 
Near the saddle-node bifurcation, the characteristic frequency $\omega^\ast$ corresponding to the peak of the spectral function 
scales as $\omega^\ast\sim |g-g_c|^{1/2}$. Furthermore, $\rho(\omega)$ behaves as $\rho(\omega<\omega^\ast) \propto \omega^{\beta_1}$ 
and $\rho(\omega>\omega^\ast) \propto \omega^{\beta_2}$, where $\beta_1-\beta_2=2$. This is quite different from the spectral 
function we obtained. One of the main reasons for this is that we derived the spectral function in the limit $\epsilon\to 0$. In this limit, 
the perturbed equation does not include the background solution explicitly. To derive such critical phenomena, 
it would be interesting to calculate the spectral function in the case $\epsilon>0$, taking into account the fluctuations of the gauge field.

{\bf Acknowledgments} 
This work was supported in part by
JSPS KAKENHI Grant Number  15K05092(AI), 23740200 (KM), and 23540326(TO).


\end{document}